\documentclass{elsart}

\usepackage{amssymb}
\usepackage{amsmath}
\usepackage{graphicx}

\newcommand{\ket}[1]{\ensuremath{|#1\rangle}}
\newcommand{\bra}[1]{\ensuremath{\langle#1|}}
\newcommand{\degree}[1]{\ensuremath{#1^\circ}}

\begin{document}

\begin{frontmatter}

\title{NMR analogues of the quantum Zeno effect}
\author{Li Xiao},
\author{Jonathan A. Jones}
\address{Centre for Quantum Computation, Clarendon Laboratory,
University of Oxford, Parks Road, Oxford OX1 3PU, United Kingdom}

\begin{abstract}
We describe Nuclear Magnetic Resonance (NMR) demonstrations of the quantum Zeno effect, and discuss briefly how these are related to similar phenomena in more
conventional NMR experiments.
\end{abstract}

\end{frontmatter}

\section*{Introduction}
It is often claimed that a watched pot never boils; in quantum mechanics this observation is in fact true, and is known as the quantum Zeno effect
\cite{misra77,pascazio96,home97}.  More specifically it is possible to suppress the coherent evolution of a quantum system by making frequent measurements which project
the quantum system onto its eigenstates.

This result is most simply described in the context of quantum information theory \cite{bennett00}.  Consider a two-level quantum system, or qubit, such as a spin-1/2
particle in a magnetic field, with two states denoted $\ket{0}$ and $\ket{1}$.  Transitions between these states can be induced, for example by the application of a
resonant electromagnetic field, causing the system to undergo coherent oscillations of the form
\begin{equation}
\ket{\psi(t)}=\cos(\omega{t}/2)\ket{0}+i\sin(\omega{t}/2)\ket{1}
\end{equation}
known as Rabi flopping, with a \textsc{not} gate, which interconverts $\ket{0}$ and $\ket{1}$, occurring when $\omega{t}=\pi$. In the language of Nuclear Magnetic
Resonance (NMR) \cite{Ernst,choreography} this corresponds to applying a $180^\circ_x$ pulse to a spin starting in the thermal equilibrium state $I_z$.

Suppose, however, that at time intervals $\tau$ a measurement is made in the $\{\ket{0},\,\ket{1}\}$ basis. The first measurement will project the system onto either
$\ket{0}$, with probability $\cos^2(\omega\tau/2)$, or $\ket{1}$, with probability $\sin^2(\omega\tau/2)$, and if the time is short, such that $\omega\tau\ll1$, the
system will almost always be found in the initial state $\ket{0}$.  Subsequent evolution and measurements will have the same effect, with the system being repeatedly
reset to the initial state.  If $n$ measurements are made at equally spaced times during a \textsc{not} gate, such that $\tau=\pi/n\omega$, then the probability that
the system will always be found in the initial state is
\begin{equation}
P_n=[\cos^2(\pi/2n)]^{n}\approx\exp(-\pi^2/4n)\approx1-\frac{\pi^2}{4n}
\end{equation}
showing that frequent measurements can effectively suppress the Rabi flopping.  This effect has been discussed in a range of quantum systems (\textit{e.g.,}
\cite{itano90}), but not as yet in NMR. We begin by showing how this can be done, and relate the Zeno effect to well-known phenomena in conventional NMR.

\section*{An one-spin NMR implementation}
NMR has provided an excellent toy system for demonstrating many quantum information phenomena \cite{jones01,vandersypen04}, but it is not immediately obvious how it can
be used to demonstrate a quantum Zeno effect. This is because the standard NMR measurement process is not the sort of projective measurement usually considered in
quantum theory, but rather a weak ensemble measurement which effectively monitors the spin system without changing it.  It is, however, perfectly possible to simulate
the effects of projective measurements in NMR by using pulsed magnetic field gradients \cite{nielsen98}.

The effect of a projective measurement on a single qubit is to project a coherent superposition of the form $\ket{\psi}=\alpha\ket{0}+\beta\ket{1}$ onto $\ket{0}$ with
probability $|\alpha|^2$ and onto $\ket{1}$ with probability $|\beta|^2$.  If the outcome of the measurement is lost, then the final state must be described by an
incoherent mixture of the form $|\alpha|^2\ket{0}\bra{0}+|\beta|^2\ket{1}\bra{1}$ showing that the key effect of measurement is just to decohere the state, removing the
off-diagonal coherence terms.  The same effect can be achieved without performing an explicit measurement by increasing the decoherence rate, that is reducing the
spin--spin relaxation time $T_2$.  In NMR this is conveniently simulated by reducing the apparent coherence time $T_2^*$ by applying a magnetic field gradient
\cite{nielsen98} which makes the Larmor frequency of the spins vary over the ensemble.  As the quantum Zeno effect only depends on the projection process, and not on
the result of the measurement, a Zeno effect should also be seen in this system.  It could be argued that gradients do not truly decohere the state, as their effects
can in principle be reversed, but the effects of diffusion within the sample mean that gradients cannot be fully reversed, and so can be indistinguishable from true
decoherence \cite{cory98}.

These ideas are easily explored in an NMR system using a single isolated spin-1/2 nucleus in exact resonance with the frequency of the applied RF. The main part of the
NMR pulse sequence comprises a series of small flip-angle RF pulses separated by delays, equivalent to a \textsc{dante} selective excitation sequence \cite{morris78}
except that gradient pulses may be applied during the delays between the RF pulses. This is followed by a gradient crush to remove any $xy$-magnetization, and a
\degree{90} observation pulse. The final NMR signal should be a single line with an intensity proportional to the remaining $I_z$ magnetization.  In the absence of the
optional gradient pulses this signal strength should show cosine oscillations, tracking the underlying Rabi flopping.  In the presence of the gradient pulses these
oscillations should be suppressed.

This sequence was applied to a standard \nuc{1}{H} NMR lineshape sample, comprising 1\% $\text{CHCl}_3$ in solution in acetone-$\text{d}_6$. All experiments were
performed on a Varian INOVA 600\,MHz spectrometer, at a temperature of \degree{20}C.  This system has a very slow spin--lattice relaxation time
($T_1\approx90\,\text{s}$), and so $T_1$ effects can be ignored during the sequence. The \nuc{1}{H} RF frequency was placed accurately on resonance, and the power
reduced so that the RF nutation rate was around \degree{1} per $\mu\text{s}$, and $1\,\mu\text{s}$ pulses were applied at 1\,ms intervals.  The number of repetitions,
$n$, was varied from 0 to 400 in steps of 10.

\begin{figure}
\begin{center}
\includegraphics[scale=0.5]{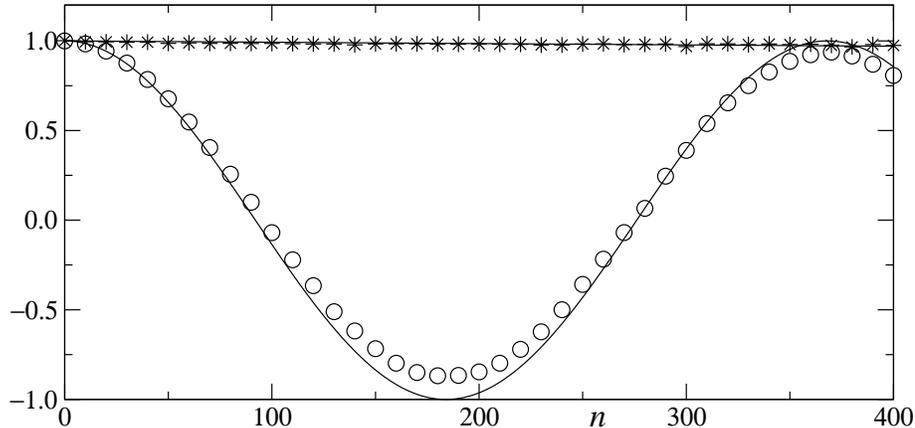}
\end{center}
\caption{Experimental results demonstrating the quantum Zeno effect in NMR.  The $x$-axis shows the number $n$ of small flip-angle (approximately $1^\circ$) pulses
applied, while the $y$-axis shows the NMR signal intensity as a fraction of the maximum intensity observed.  Stars and circles show data points with and without the
application of gradient pulses between the RF pulses. The smooth lines show a cosine oscillation $\cos(n\theta)$ and an exponential decay $\exp(-kn)$, with fitted
values $\theta=0.978^\circ$ and $k=7.5\times10^{-5}$.}\label{fig:data}
\end{figure}
The results are plotted in Figure~\ref{fig:data}, and show the form expected.  In the absence of gradients the observed $I_z$ magnetization undergoes cosine modulation,
with a period of $n\approx360$, while in the presence of gradients this modulation is almost completely suppressed.  The deviations from perfect cosine modulation can
be ascribed to a combination of $B_1$ inhomogeneity and $T_2^*$ decoherence during the \textsc{dante} sequence. The slight exponential decay visible in the data
acquired with gradients is consistent with the Zeno effect, and occurs because the interval between the measurements is not quite zero.

\section*{Discussion}
The experiment above corresponds to the simplest version of the Zeno effect, in which measurements can be treated as occurring instantaneously at certain points in the
evolution, but it is also possible to relax this limit slightly.  As long as the measurement occurs very rapidly in comparison with the evolution rate, the overall
effect will be similar (although some authors do not consider this to be a true Zeno effect \cite{home97}). When measurement is replaced by gradient-induced decoherence
the timescale of the ``measurement'' is inversely proportional to the strength of the gradient field, and so a Zeno-like effect can be seen by attempting to excite a
sample with an RF field \textit{during} the application of a field gradient. Clearly this will be ineffective if the gradient is strong compared with the RF field, that
is the measurement is fast compared with the nutation frequency.  Indeed, as there is no real difference between frequency variation arising from gradients and that
arising from intrinsic interactions, the frequency-selective behaviour of a \textsc{dante} sequence can be seen as an example of the Zeno effect.  In the same way, the
effect of a spin-lock field in suppressing Zeeman evolution can be seen as Zeno-like. A more thorough discussion of the relationship between quantum Zeno effects and
decoupling sequences can be found elsewhere \cite{facchi04}.

\begin{figure}
\begin{center}
\includegraphics[scale=1]{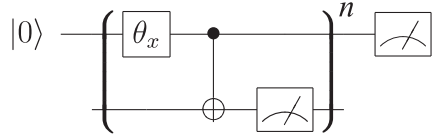}
\end{center}
\caption{A quantum circuit for exploring the Zeno effect in a two qubit system.  The section enclosed in parentheses is repeated $n$ times.}\label{fig:twoqubit}
\end{figure}

Despite the comments above, this one-spin demonstration is perhaps \textit{too} simple to be of any real interest.  In particular it is well known that the dynamics of
a single isolated spin is indistinguishable from that of a classical magnetic moment, and so the results above can be understood using the classical vector model
\cite{choreography}.  To address this we consider a two-qubit system in which something closer to traditional quantum measurements can be performed. The
controlled-\textsc{not} gate can be considered as a measurement gate, as it causes the target qubit to become correlated with the control qubit, and decoherence of the
target qubit (whether spontaneous or artificially induced) completes the measurement process.  A suitable quantum circuit is shown in figure~\ref{fig:twoqubit} where
the section enclosed in parentheses is repeated $n$ times. Note that it is not necessary to know the results of the measurements on the target qubit, as these are just
discarded, and so it is not necessary to initialize this qubit before the start of the circuit; indeed it is simplest to assume that this qubit starts in the maximally
mixed state. The final result is obtained by measuring the control qubit in the computational basis.

\section*{A two-spin NMR implementation}
This circuit can be easily implemented using NMR techniques in a heteronuclear two-spin system.  The controlled-\textsc{not} gate is normally constructed from
pseudo-Hadamard gates \cite{jones01}, single-qubit $z$-rotations (usually implemented using frame rotations \cite{knill00}), and periods of evolution under the
J-coupling between the two spins.  For our experiments this approach can be simplified in two ways.  Firstly the fact that the initial state of the target qubit is
irrelevant means that the initial pseudo-Hadamard gate can be dropped.  Secondly the RF reference frequencies for the two spins can be chosen to be in resonance with
the high-frequency components of the two multiplets, such that free evolution implements \textit{both} the desired J-coupling and the $z$-rotations simultaneously.  The
selective decoherence of the target spin can be implemented by applying two magnetic field gradients with equal but opposite strengths separated by a $180^\circ$ pulse
applied to the target spin.  This pulse will also refocus the natural evolution of the two-spin system except for the chemical shift evolution of the control spin; this
can be refocused stroboscopically, that is by choosing the total length of the decoherence period as an integer multiple of the inverse of the evolution frequency.

\begin{figure}
\begin{center}
\includegraphics[scale=1]{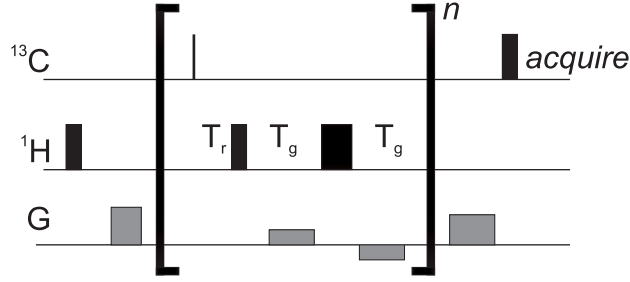}
\end{center}
\caption{The NMR pulse sequence used to implement the quantum circuit shown in Figure~\ref{fig:twoqubit} in a spin system comprising a \nuc{1}{H} and a \nuc{13}{C}
nucleus. The section enclosed in parentheses is repeated $n$ times. RF pulses are shown in black, with thin, medium and thick boxes indicating flip angles of
\degree{5}, \degree{90} and \degree{180} respectively.  Gradient pulses are shown as grey boxes on the line labeled G, with height indicating the strength of each
pulse. The strength of the measurement is determined by setting the delay $T_r=1/2rJ$, where $J$ is the scalar coupling constant, to obtain the $r^\text{th}$ root of
controlled-\textsc{not} gate. The period $T_g$, during which gradients are applied, is chosen stroboscopically as $1/J$.}\label{fig:twoqubitpulses}
\end{figure}

This approach can be easily extended to study weak measurements, by replacing the controlled-\textsc{not} gate with its $r^\text{th}$ root, that is a gate which applies
a $(\degree{180}/r)_x$ rotation to the target qubit if and only if the control qubit is in the state \ket{1}, which can be implemented in a similar way to a
controlled-\textsc{not} gate \cite{jones01}.  The final NMR pulse sequence is shown in figure~\ref{fig:twoqubitpulses}.  We chose to use the spin system provided by the
\nuc{1}{H} and \nuc{13}{C} nuclei in a sample of 10\,mg of \nuc{13}{C} labeled sodium formate ($\text{Na}^+\text{HCO}_2^-$) dissolved in 0.75\,ml of
$\text{D}_2\text{O}$ at a temperature of $20^\circ\text{C}$ \cite{xiao05}. The effects of relaxation are much more important in this system, both because of the shorter
relaxation times and because of the length of the measurement (imposed by the requirement to work stroboscopically).  For this reason the flip angle $\theta$ of the
weak pulses was increased from around \degree{1} to \degree{5}. This larger flip angle means that the Zeno effect will not be as effective as in the one qubit system.

With this more complex pulse sequence it is necessary to think more carefully about the effects of the gradients than was the case with a single qubit.  In particular
it appears that the negative gradient will act to partly cancel the positive gradient applied during the following iteration \cite{choreography}.  A full analysis shows
that the situation is more complex than this, but this is a genuine concern.  The problem is further complicated by the presence of diffusion, which acts to make
successive gradients partially independent of one another. This helps resolve the problem described above, but can introduce a new problem, as in the presence of strong
diffusion the direct effects of the gradients on the control spin will not be refocused.  We addressed this compound problem in two ways.  Firstly, our gradient pulses
do not last for the whole period $T_g$ (see figure~\ref{fig:twoqubitpulses}), but are somewhat shorter and are placed tightly around the \degree{180} pulse. This acts
to maximise the effects of diffusion \textit{between} measurements, while minimising them \textit{within} measurements.  Secondly, we chose to use the \nuc{13}{C} as
our control qubit, with the \nuc{1}{H} spin as the target qubit. As the size of diffusive effects depends on the square of the gyromagnetic ratio \cite{cory98} these
effects will be 16 times larger for the \nuc{1}{H} target spin (where we want them to be large) than for the \nuc{13}{C} control spin (where we want them to be small).

\begin{figure}
\begin{center}
\includegraphics[scale=0.5]{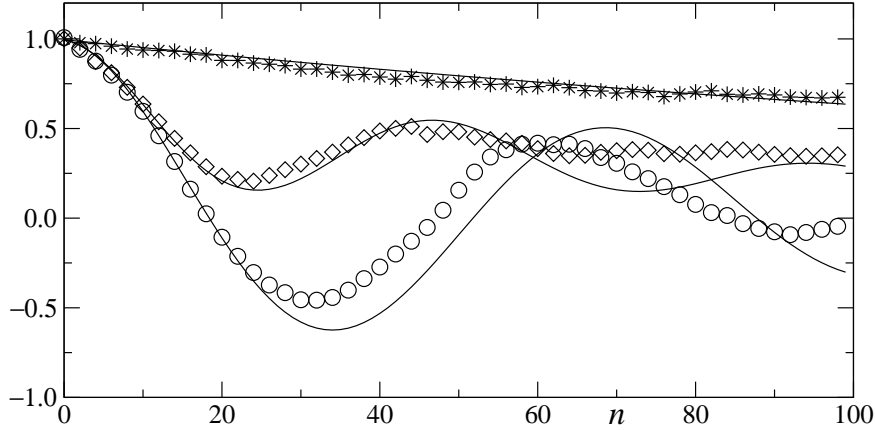}
\end{center}
\caption{Experimental results demonstrating the quantum Zeno effect in a two spin NMR system. The $x$-axis shows the number $n$ of small flip-angle (approximately
$5^\circ$) pulses applied, while the $y$-axis shows the NMR signal intensity as a fraction of the maximum intensity observed.  Stars show data points with strong
measurement ($r=1$) while diamonds and circles indicate relatively weak measurements ($r=16$ and $r=64$ respectively).  The smooth lines show calculated decays.
}\label{fig:twoqubitdata}
\end{figure}

The results of implementing this sequence are shown in figure~\ref{fig:twoqubitdata}.  Data is shown for strong measurements ($r=1$),  weak measurements ($r=64$) and
intermediate measurements ($r=16$), and a calculated line, obtained by numerical simulation, is plotted in each case.  As expected, strong measurements are effective in
suppressing the Rabi oscillations, while weaker measurements have less effect.  (The Zeno effect is not as clear in this plot as in Figure 1 because of the larger flip
angle used between measurements.) There is broad agreement between experimental and calculated results, but deviations can be seen. These can be largely explained by
considering imperfections in the gradients used to implement measurements on the target spin, in particular the effects of diffusion.  Firstly diffusion means that the
dephasing of the control spin is not perfectly refocused, and so a weak ``background'' measurement occurs even in the case of very large values of $r$. Secondly,
interactions between gradients on successive measurements cannot be completely ignored.  These imperfections mean that strong measurements will not be perfectly strong,
and weak measurements will not be perfectly weak, leading to the deviations observed.

\section*{Conclusions}
The Zeno effect in NMR experiments has not previously been explored, probably as a result of the difficulty of performing true quantum measurements.  It is, however,
possible to implement effective measurements using field gradients, allowing the effect to be easily demonstrated.  Drawing an analogy between gradients and naturally
occurring variations in the Larmor frequency of spins leads to a link between the Zeno effect with weak measurements and the behaviour of frequency selective pulses.
This can be explored in more detail using a two spin system where the measurement strength can be easily controlled, and the expected results are seen.

\section*{Acknowledgements}
We thank the UK EPSRC and BBSRC for financial support.


\begin{thebibliography}{00}

\bibitem{misra77}
B.~Misra, E.~C.~G. Sudarshan,
J.~Math. Phys. 18 (1977) 756.

\bibitem{pascazio96}
H.~Nakazato, M.~Namiki, S.~Pascazio, H.~Rauch,
Phys. Lett. A 217 (1996) 203.

\bibitem{home97}
D.~Home, M.~A.~B. Whitaker,
Ann. Phys. (N.~Y.) 258 (1997) 237.

\bibitem{bennett00}
C.~H. Bennett, D.~P. DiVincenzo,
Nature (Lond.) 404 (2000) 247

\bibitem{Ernst}
R.~R. Ernst, G.~Bodenhaause and A.~Wokaun, Principles of Nuclear Magnetic Resonance in One and Two Dimensions, Clarendon Press, Oxford, 1987.

\bibitem{choreography}
R.~Freeman, Spin Choreography, Spektrum, Oxford, 1997.

\bibitem{itano90}
W.~M. Itano, D.~J. Heinzen, J.~J. Bollinger, D.~J. Wineland,
Phys. Rev. A 41 (1990), 2295.

\bibitem{jones01}
J.~A. Jones,
Prog. NMR Spectrosc. 38 (2001) 325.

\bibitem{vandersypen04}
L.~M.~K. Vandersypen, I.~L. Chuang,
Rev. Mod. Phys. 76 (2004) 1037.

\bibitem{nielsen98}
M.~A. Nielsen, E.~Knill, R.~Laflamme,
Nature (Lond.) 396(1998) 52.

\bibitem{cory98}
D.~G. Cory, M.~D. Price, W.~Maas, E.~Knill, R.~Laflamme, W.~H. Zurek, T.~F. Havel, S.~S. Somaroo,
Phys. Rev. Lett. 81 (1998) 2152.

\bibitem{morris78}
G.~A. Morris, R.~Freeman,
J. Magn. Reson. 29 (1978) 433.


\bibitem{facchi04}
P.~Facchi, D.~A. Lidar, S.~Pascazio,
Phys. Rev. A 69 (2004) 032314.

\bibitem{knill00}
E.~Knill, R.~Laflamme, R.~Martinez, C.-H. Tseng, Nature 404 (2000)
368.

\bibitem{xiao05}
L.~Xiao. J.~A. Jones, Phys. Rev. A 72 (2005) 032326.

\end{thebibliography}
\end{document}